\documentclass[12pt]{article}

\begin{document}


\vskip 1cm

\begin{center}
{\Large {\bf Three-loop universal anomalous dimension of the Wilson operators
in ${\mathcal N}=4$ Supersimmetric Yang-Mills theory}\footnote{This work was supported by
the RFBR grant 04-02-17094,
RSGSS-1124.2003.2}} \\[0pt]
\vspace{1.5cm} {\large \ A.~V.~Kotikov$^a$, L.~N.~Lipatov$^b$, A.~I.~Onishchenko$^{b,c}$ and V.~N.~Velizhanin$^b$ } \\[0pt]
\vspace{1cm} {$^a$ \em Bogoliubov Laboratory of Theoretical Physics \\[0pt]
Joint Institute for Nuclear Research\\[0pt]
141980 Dubna, Russia }\\[0pt]
\vspace{0.5cm} {$^b$ \em Theoretical Physics Department\\[0pt]
Petersburg Nuclear Physics Institute\\[0pt]
Orlova Roscha, Gatchina\\[0pt]
188300, St. Petersburg, Russia }\\[0pt]
\vspace{0.5cm} {$^c$ \em II. Institut f\" ur Theoretische Physik, Universit\" at Hamburg\\[0pt]
Luruper Chaussee 149, 22761 Hamburg, Germany }\\[0pt]
\end{center}

\vspace{1.5cm}\noindent

\abstract{
\noindent We present results for the three-loop universal anomalous dimension $\gamma _{uni}(j)\,$
of Wilson twist-2 operators in the ${\mathcal N}=4$ Supersymmetric Yang-Mills theory.
These expressions are obtained by extracting the most complicated contributions
from the three-loop anomalous dimensions in QCD.
This result is in an agreement with the hypothesis of the integrability
of ${\mathcal N}=4$ Supersymmetric Yang-Mills theory in the context of AdS/CFT-correspondence.}

\section{Introduction}

The anomalous dimensions of the twist-2 Wilson operators govern the
Bjorken scaling violation for parton distributions in a framework of Quantum
Chromodynamics (QCD). These quantities are
given by the Mellin transformation
\begin{equation}
\gamma _{ab}(j)=\int_{0}^{1}dx\,\,x^{j-1}W_{b\rightarrow a}(x)
\end{equation}
of the splitting kernels $W_{b\rightarrow a}(x)$ for the
Dokshitzer-Gribov-Lipatov-Altarelli-Parisi (DGLAP) equation~\cite{DGLAP}
which evolves the parton densities $f_{a}(x,Q^{2})$ (hereafter $a=\lambda,\,g,\,\phi$ for
the spinor, vector and scalar particles, respectively)
as follows
\begin{equation}
\frac{d}{d\ln {Q^{2}}}f_{a}(x,Q^{2})=\int_{x}^{1}\frac{dy}{y}
\sum_{b}W_{b\rightarrow a}(x/y)\,f_{b}(y,Q^{2})\,.
\end{equation}
The anomalous dimensions and splitting kernels in QCD are known up to the
next-to-next-to-leading order (NNLO) of the perturbation theory~\cite{LONLOAD,VMV}.

The QCD expressions for anomalous dimensions can be transformed to the case of the
${\mathcal N}$-extended Supersymmetric Yang-Mills theories (SYM)
if one will use for the Casimir operators $C_{A},C_{F},T_{f}$ the following
values $C_{A}=C_{F}=N_{c}$, $T_{f}n_f={\mathcal N}N_{c}/2$.
For ${\mathcal N}\!\!=\!\!2$ and ${\mathcal N}\!\!=\!\!4$-extended SYM the anomalous dimensions of the Wilson operators get also
additional contributions coming from scalar particles~\cite{KL}.
These anomalous dimensions were calculated in the next-to-leading order (NLO)~\cite{KoLiVe}
for the ${\mathcal N}=4$ SYM.

However, it turns out, that the expressions
for eigenvalues of the anomalous dimension matrix in the ${\mathcal N}=4$ SYM can be derived directly
from the QCD anomalous dimensions without tedious calculations by using a
number of plausible arguments. The method elaborated in Ref.~\cite{KL} for
this purpose is based on special properties of
the integral kernel for
the Balitsky-Fadin-Kuraev-Lipatov (BFKL) equation~\cite{BFKL,next} in this model
and a new relation between the BFKL and DGLAP equations (see~\cite{KL00}).
In the NLO approximation this method gives the correct results for
anomalous dimensions eigenvalues, which was checked by {\it direct calculations} in Ref.~\cite{KoLiVe}.
Its properties will be reviewed below only shortly and
a more extended discussion can be found in~\cite{KL}.
Using the results for the NNLO
corrections to anomalous dimensions in QCD~\cite{VMV}
and the method of Ref.~\cite{KL} we derive the
eigenvalues of the anomalous dimension matrix for the ${\mathcal N}=4$ SYM in the NNLO
approximation~\cite{KLOV}.

The obtained result is very important for the verification of the
various assumptions~\cite{Arutyunov:2001mh,Beisert:2003tq,Eden:2004ua,
Staudacher:2004tk}
coming from the investigations
of the properties of a conformal operators in the context of AdS/CFT correspondence~\cite{AdS-CFT}.

\section{Evolution equation in ${\mathcal N}=4$ SYM}

The reason to investigate the BFKL and DGLAP equations in the case of
supersymmetric theories is
related to
a common belief, that the high symmetry
may significantly simplify their structure. Indeed, it was
found in the leading logarithmic approximation (LLA)~\cite{BFKL2}, that the
so-called quasi-partonic operators in ${\mathcal N}=1$ SYM are unified in
supermultiplets with anomalous dimensions obtained from the universal
anomalous dimension $\gamma_{uni}(j)$ by shifting its argument by an
integer number. Further, the anomalous dimension matrices for twist-2
operators are fixed by the superconformal invariance~\cite{BFKL2}.
Calculations in the maximally extended ${\mathcal N}=4$ SYM, where the
coupling constant is not renormalized, give even more remarkable
results. Namely, it turns out, that here all twist-2 operators enter in
the same multiplet, their anomalous dimension matrix is fixed completely
by the super-conformal invariance  and its universal anomalous dimension
in LLA is proportional to $\Psi (j-1)-\Psi (1)$, which means, that the
evolution equations for the matrix elements of quasi-partonic operators in
the multicolour limit $N_{c}\rightarrow \infty $ are equivalent to
the Schr\"{o}dinger equation for
an integrable Heisenberg spin model~\cite{N=4,LN4}. In QCD the
integrability remains only in a small sector of these operators~\cite{BDMB} (see also~\cite{Ferretti:2004ba}). In the case of ${\mathcal N}=4$ SYM
the equations for other sets of operators are also
integrable~\cite{Minahan:2002ve,Beisert:2003tq,Beisert:2003yb}.

Similar results related
to the integrability of the multi-colour QCD were obtained
earlier in the Regge limit~\cite{Integr}. Moreover, it was shown~\cite{KL00},
that in the ${\mathcal N}=4$ SYM there is a deep relation between the BFKL and DGLAP
evolution equations. Namely, the $j$-plane singularities of anomalous dimensions of the Wilson
twist-2 operators in this case can be obtained from the eigenvalues of the
BFKL kernel by their analytic continuation. The NLO
calculations in ${\mathcal N}=4$ SYM demonstrated~\cite{KL}, that some of these
relations are valid also in higher orders of perturbation theory. In
particular, the BFKL equation has the property of the hermitian
separability, the linear combinations of the multiplicatively renormalized
operators do not depend on the coupling constant, the eigenvalues of the
anomalous dimension matrix are expressed in terms of the universal
function $\gamma _{uni}(j)$ which can be obtained also from the BFKL
equation~\cite{KL}. The results for  $\gamma _{uni}(j)$ were checked
by {\it direct calculations} in Ref.~\cite{KoLiVe}

\section{Method of obtaining the eigenvalues of the AD matrix in ${\mathcal N}=4$ SYM}\label{MethodAD}

In the ${\mathcal N}=4$ SYM theory~\cite{BSSGSO}
one can introduce the following colour and $SU(4)$ singlet local Wilson twist-2
operators~\cite{KL,KoLiVe}:
\begin{eqnarray}
\mathcal{O}_{\mu _{1},...,\mu _{j}}^{g} &=&\hat{S}
G_{\rho \mu_{1}}^{a}{\mathcal D}_{\mu _{2}}
{\mathcal D}_{\mu _{3}}...{\mathcal D}_{\mu _{j-1}}G_{\rho \mu _{j}}^a\,,
\label{ggs}\\
{\tilde{\mathcal{O}}}_{\mu _{1},...,\mu _{j}}^{g} &=&\hat{S}
G_{\rho \mu_{1}}^a {\mathcal D}_{\mu _{2}}
{\mathcal D}_{\mu _{3}}...{\mathcal D}_{\mu _{j-1}}{\tilde{G}}_{\rho \mu _{j}}^a\,,
\label{ggp}\\
\mathcal{O}_{\mu _{1},...,\mu _{j}}^{\lambda } &=&\hat{S}
\bar{\lambda}_{i}^{a}\gamma _{\mu _{1}}
{\mathcal D}_{\mu _{2}}...{\mathcal D}_{\mu _{j}}\lambda ^{a\;i}\,, \label{qqs}\\
{\tilde{\mathcal{O}}}_{\mu _{1},...,\mu _{j}}^{\lambda } &=&\hat{S}
\bar{\lambda}_{i}^{a}\gamma _{5}\gamma _{\mu _{1}}{\mathcal D}_{\mu _{2}}...
{\mathcal D}_{\mu_{j}}\lambda ^{a\;i}\,, \label{qqp}\\
\mathcal{O}_{\mu _{1},...,\mu _{j}}^{\phi } &=&\hat{S}
\bar{\phi}_{r}^{a}{\mathcal D}_{\mu _{1}}
{\mathcal D}_{\mu _{2}}...{\mathcal D}_{\mu _{j}}\phi _{r}^{a}\,,\label{phphs}
\end{eqnarray}
where ${\mathcal D}_{\mu }$ are covariant derivatives.
The spinors $\lambda _{i}$ and
field tensor $G_{\rho \mu }$ describe gluinos and gluons, respectively, and
$\phi _{r}$ are the complex scalar fields.
For all operators in Eqs.~(\ref{ggs})-(\ref{phphs}) the symmetrization of the tensors
in the Lorentz indices
$\mu_{1},...,\mu _{j}$ and a subtraction of their traces is assumed.
Due to the fact that all twist-2 operators belong to the same supermultiplet
the eigenvalues of anomalous dimensions matrix can be expressed through
one universal anomalous dimension $\gamma_{uni}(j)$ with shifted
argument\footnote{Non-diagonal elements of the anomalous dimensions matrix
are related with non-forward anomalous dimensions by means of superconformal
Ward identities~\cite{SUSYCWI} and can be expressed also through
{\it non-forward} universal anomalous dimension~\cite{OVJHEP}.}.

As it was already pointed out in the Introduction, the universal anomalous
dimension can be extracted directly from the QCD results without finding the
scalar particle contribution. This possibility is based on the deep
relation between the DGLAP and BFKL dynamics in the ${\mathcal N}=4$ SYM~
\cite{KL00,KL}.

To begin with, the eigenvalues of the BFKL kernel turn out to be
analytic functions of the conformal spin $\left| n\right| $ at least in two first orders of
perturbation theory \cite{KL}. Further, in the framework of the ${\overline{\mathrm{DR}}}$-scheme~\cite{DRED}
one can obtain from the BFKL equation (see~\cite{KL00}), that there is no
mixing among the special functions of different transcendentality levels $i$
\footnote{
Note that similar arguments were used also in~\cite{FleKoVe} to obtain
analytic results for contributions of some complicated massive Feynman
diagrams without direct calculations.},
i.e. all special functions at the NLO correction contain only sums of the
terms $\sim 1/j^{i}~(i=3)$. More precisely, if we introduce the
transcendentality level $i$ for the eigenvalues $\omega(\gamma)$ of integral kernels of the BFKL
equations
in an accordance with the complexity of the terms in the
corresponding sums
\[
\Psi \sim 1/\gamma ,~~~\Psi ^{\prime }\sim \beta ^{\prime }\sim \zeta
(2)\sim 1/\gamma ^{2},~~~\Psi ^{\prime \prime }\sim \beta ^{\prime \prime
}\sim \zeta (3)\sim 1/\gamma ^{3},
\]
then for the BFKL kernel in the leading order (LO) and in NLO the
corresponding levels are $i=1$ and $i=3$, respectively.

Because in ${\mathcal N}=4$ SYM there is a relation between the BFKL and DGLAP equations
(see~\cite{KL00,KL}), the similar properties should be valid for the
anomalous dimensions themselves, i.e. the basic functions $\gamma
_{uni}^{(0)}(j)$, $\gamma _{uni}^{(1)}(j)$ and $\gamma _{uni}^{(2)}(j)$ are
assumed to be of the types $\sim 1/j^{i}$ with the levels $i=1$, $i=3$ and
$i=5$, respectively. An exception could be for the terms appearing at a given
order from previous orders of the perturbation theory. Such
contributions could be generated and/or removed by an approximate finite
renormalization of the coupling constant. But these terms do not appear in
the ${\overline{\mathrm{DR}}}$-scheme.

It is known, that at the LO and NLO approximations
(with the SUSY relation for the QCD color factors $C_{F}=C_{A}=N_{c}$) the
most complicated contributions (with $i=1$ and $i=3$, respectively) are the
same for all LO and NLO anomalous dimensions in QCD~\cite{LONLOAD}
and for the LO and NLO scalar-scalar anomalous
dimensions~\cite{KoLiVe}. This property allows one to find the
universal anomalous dimensions $\gamma _{uni}^{(0)}(j)$ and $\gamma
_{uni}^{(1)}(j)$ without knowing all elements of the anomalous dimensions
matrix~\cite{KL}, which was verified by the exact calculations in~\cite{KoLiVe}.

Using above arguments, we conclude, that at the NNLO level there is only one
possible candidate for $\gamma _{uni}^{(2)}(j)$. Namely, it is the most
complicated part of the QCD anomalous dimensions matrix
(with
the SUSY relation for the QCD color factors
$C_{F}=C_{A}=N_{c}$).
Indeed, after the diagonalization of the
anomalous dimensions matrix its eigenvalues should have this most complicated part
as a common contribution because they differ each from others only by a shift of
the argument and their differences are constructed from
less complicated terms. The non-diagonal matrix elements of the anomalous dimensions matrix
contain also only less complicated terms (see, for example, anomalous dimensions exact
expressions at LO and NLO approximations
in Refs.~\cite{LONLOAD}
for QCD
and~\cite{KoLiVe} for ${\mathcal N}=4$ SYM) and therefore they cannot generate
the most complicated contributions to the eigenvalues of anomalous dimensions matrix.

Thus, the most complicated part of the NNLO QCD
anomalous dimensions should coincide (up to color factors)
with the universal anomalous dimension $\gamma_{uni}^{(2)}(j)$.

\section{NNLO anomalous dimension for ${\mathcal N}=4$ SYM}

The final three-loop result
\footnote{
Note, that in an accordance with Ref.~\cite{next}
 our normalization of $\gamma (j)$ contains
the extra factor $-1/2$ in comparison with
the standard normalization (see~\cite{LONLOAD})
and differs by sign in comparison with one from Ref.~\cite{VMV}.}
for the universal anomalous dimension $\gamma_{uni}(j)$
for ${\mathcal N}=4$ SYM is~\cite{KLOV}
\begin{eqnarray}
\gamma(j)\equiv\gamma_{uni}(j) ~=~ \hat a \gamma^{(0)}_{uni}(j)+\hat a^2
\gamma^{(1)}_{uni}(j) +\hat a^3 \gamma^{(2)}_{uni}(j) + ... , \qquad \hat a=\frac{\alpha N_c}{4\pi}\,,  \label{uni1}
\end{eqnarray}
where\footnote{Note, that
$\gamma^{(1)}_{uni}(j)$ was obtained also by {\it direct calculations} in
Ref.~\cite{KoLiVe}.}
\begin{eqnarray}
\frac{1}{4} \, \gamma^{(0)}_{uni}(j+2) &=& - S_1,  \label{uni1.1} \\
\frac{1}{8} \, \gamma^{(1)}_{uni}(j+2) &=& \Bigl(S_{3} + \overline S_{-3} \Bigr) -
2\,\overline S_{-2,1} + 2\,S_1\Bigl(S_{2} + \overline S_{-2} \Bigr),  \label{uni1.2} \\
\frac{1}{32} \, \gamma^{(2)}_{uni}(j+2) &=& 2\,\overline S_{-3}\,S_2 -S_5 -
2\,\overline S_{-2}\,S_3 - 3\,\overline S_{-5}  +24\,\overline S_{-2,1,1,1}\nonumber\\
&&\hspace{-1.5cm}+ 6\biggl(\overline S_{-4,1} + \overline S_{-3,2} + \overline S_{-2,3}\biggr)
- 12\biggl(\overline S_{-3,1,1} + \overline S_{-2,1,2} + \overline S_{-2,2,1}\biggr)\nonumber \\
&& \hspace{-1.5cm}  -
\biggl(S_2 + 2\,S_1^2\biggr) \biggl( 3 \,\overline S_{-3} + S_3 - 2\, \overline S_{-2,1}\biggr)
- S_1\biggl(8\,\overline S_{-4} + \overline S_{-2}^2\nonumber \\
&& \hspace{-1.5cm}  + 4\,S_2\,\overline S_{-2} +
2\,S_2^2 + 3\,S_4 - 12\, \overline S_{-3,1} - 10\, \overline S_{-2,2} + 16\, \overline S_{-2,1,1}\biggr)
\label{uni1.5}
\end{eqnarray}
and $S_{a} \equiv S_{a}(j),\ S_{a,b} \equiv S_{a,b}(j),\ S_{a,b,c} \equiv
S_{a,b,c}(j)$ are harmonic sums
\begin{eqnarray}
&&\hspace*{-1cm} S_{a}(j)\ =\ \sum^j_{m=1} \frac{1}{m^a},
\ \ S_{a,b,c,\cdots}(j)~=~ \sum^j_{m=1}
\frac{1}{m^a}\, S_{b,c,\cdots}(m),  \label{ha1} \\
&&\hspace*{-1cm} S_{-a}(j)~=~ \sum^j_{m=1} \frac{(-1)^m}{m^a},~~
S_{-a,b,c,\cdots}(j)~=~ \sum^j_{m=1} \frac{(-1)^m}{m^a}\,
S_{b,c,\cdots}(m),  \nonumber \\
&&\hspace*{-1cm} \overline S_{-a,b,c,\cdots}(j) ~=~ (-1)^j \, S_{-a,b,c,...}(j)
+ S_{-a,b,c,\cdots}(\infty) \, \Bigl( 1-(-1)^j \Bigr).  \label{ha3}
\end{eqnarray}

The expression~(\ref{ha3}) is defined for all integer values of arguments
(see~\cite{KK,KL,KoVe})
but can be easily analytically continued to real and complex $j$
by the method of Refs.~\cite{AnalCont,KL,KoVe}.

\section{Integrability and the AdS/CFT-correspondence}

The investigation of the integrability in ${\mathcal N}=4$ SYM for
a BMN-operators~\cite{Berenstein:2002jq} gives a possibility to find the anomalous
dimension of a Konishi operators~\cite{Beisert:2003tq}, which has the anomalous dimension
coinciding with our expression~ (\ref{uni1}) for $j=4$
\begin{equation}\label{ADKonTL}
\gamma_{uni}(j)\big|_{j=4}
=-6\, \hat a + 24\, \hat a^2 - 168\,\hat a^3=
-\frac{3\,\alpha\,N_c}{2\pi}
+\frac{3\,\alpha^2\,N_c^2}{2\pi^2}
-\frac{21\,\alpha^3\,N_c^3}{8\pi^4}\,.
\end{equation}
It is confirmed also by direct calculation in two~\cite{Arutyunov:2001mh,KoLiVe} and three-loop~\cite{Eden:2004ua}
orders.

A very interesting result comes from the consideration of the factorized S-matrix~\cite{Staudacher:2004tk},
which based on the investigation of the both side of AdS/CFT-correspondence
\cite{Berenstein:2002jq,Minahan:2002ve,Beisert:2003tq,Beisert:2003yb,
Beisert:2003ys,Arutyunov:2003uj}
and gives a possibility to find three-loop anomalous dimension from the Bethe
ansatz for arbitrary values of the Lorenz spin. The resulting Bethe ansatz reproduces our results
for universal anomalous dimension $\gamma_{uni}(j)$ Eq.~(\ref{uni1}) and, then, confirm the hypotheses on integrability
in ${\mathcal N}=4$ SYM.

\section{Conclusion}

We found the NNLO anomalous dimension $\gamma _{uni}(j)$ in the ${\mathcal N}=4$ SYM~\cite{KLOV}.
This universal anomalous dimension at $j=4$ was used to calculate the anomalous dimension
of Konishi operator up to 3-loops. It is remarkable, that our results coincide
with corresponding expression obtained from dilatation operator approach
and integrability \cite{Beisert:2003tq,Beisert:2003ys}.
Moreover, these results for the universal anomalous dimension was used for a verification
of the S-matrix approach to AdS/CFT-correspondence~\cite{Staudacher:2004tk},
which is based on the integrability of the corresponding dual theories at large-$N_C$ limit.

\end{document}